\documentclass[aps,prl,twocolumn,prl]{revtex4-1}

\usepackage{setspace}
\usepackage{amsmath}
\usepackage{graphicx}

\usepackage{caption}
\usepackage{subcaption}

\usepackage{hyperref}

\begin{document}
\pdfoutput=1
%\preprint{}

\title{Friction between Ring Polymer Brushes}
\author{Aykut Erbas}
\affiliation{Northwestern University, Department of Materials Science and Engineering\\ Evanston, IL 60208, USA}
\author{Jaros{\l}aw Paturej}
\affiliation{Leibniz-Institut of Polymer Research, 01069 Dresden, Germany}
\affiliation{Institute of Physics, University of Szczecin, Wielkopolska 15,
70451 Szczecin, Poland}

%\author{}
%\thanks{}
%\affiliation{}


\date{\today}

\begin{abstract}

Friction between ring-polymer brushes at melt densities sliding past each other are studied using extensive course-grained molecular dynamics simulations and scaling arguments, and the results are compared  to the friction between linear-polymer brushes. We show that for a velocity range spanning over three decades, the frictional forces measured for ring-polymer brushes  are  half the corresponding friction in case of linear brushes. In the linear-force regime,  the weak inter-digitation of two ring brushes compared to linear brushes also leads to a lower number of binary collisions between the monomers of opposing brushes.  At high velocities, where the thickness of the inter-digitation layer between two opposing brushes is on the order monomer size regardless of brush topology,  stretched segments of ring polymers take a double-stranded conformation.  As a result,  monomers of the double-stranded segments collide less with the monomers of the opposing ring brush even though a similar number of monomers occupies the inter-digitation layer for ring and linear-brush bilayers. The numerical data obtained from our simulations is consistent with the proposed scaling analysis. Conformation-dependent frictional reduction observed  in ring brushes can have important consequences in non-equilibrium bulk systems.

\end{abstract}

%\keywords{}

\maketitle

\section{Introduction}

If  polymer chains are grafted by one of their ends to a planar or curved surface, above  a certain critical grafting density $\rho_g^* \approx 1/R_0^2$~\cite{DeGennes:1987ts}, where $R_0$ is the characteristic equilibrium chain size,  the chains are stretched away from the surface due to steric repulsion from surrounding chains and form polymer brushes.
A polymer brush is a soft polymeric material that can deform under various external forces. The external force can be due to a fluid flowing over the brush, a  flow due to the relative motion of a second brush, or alternatively, an external electrical field (if the polymers are charged).  Once the external stimulus is completely removed, as the chains constituting the brush relax, the deformation of brush is reversed similar to the elastic deformation observed for solids. However,  the tribological behaviour of polymeric systems more resembles that of a fluid rather than a solid~\cite{Gong:2006ks}.  For instance, if two  inter-digitated polymer brushes are slid past each other, frictional forces due to the relative motion of brushes vanish linearly as the relative velocity of the brushes is decreased towards zero. In other words, the friction is viscous, for which no static friction occurs, unlike the solid-state friction, where finite forces are needed to initiate motion~\cite{persson}.

The polymer brushes has attracted increasing attention due to their applications in nanotechnology and material sciences as bio-sensors, bio-fueling, stimuli-responsive surfaces~\cite{Ayres:2010by,Halperin:1999ic,Zhou:2006db,Stuart:2019hu,Sidorenko:1999bn}, or for the stabilisation of colloidal solutions~\cite{Pincus:1991wu,Currie:2003wf}. The most fascinating application of brush-like structures is facilitated by nature in maintaining the lubrication in tissues~\cite{Raviv:2003jb,Urbakh:2004wr,Klein:1994ux,Klein:2009wj}. For instance, in mammalian joints, where very low lubrication should be retained under pressures as high as 5 MPa, brush-like structures in combination with the synovial fluid provide lubrication in between the articular joints~\cite{Chen:2009gs,Klein:2009wj,Zappone:2007ev}. The surfaces separating the articular cartilage and the synovial fluid are thought to be covered by high molecular weight molecules such as lubricin~\cite{Klein:2009wj, Benz:2004kl}. In turn, these long and charged glycoproteins  may function as water-based bio-lubricants.  The relationship between  morphology of these long and charged macromolecules on the cartilage surfaces and their function in lubrication is still under investigation~\cite{Maeda:2002uz,Zappone:2007ev,Klein:2009wj,Benz:2004kl,Raviv:2003jb,Chang:2008hb,Banquy:2014dr}.
Amphiphilic lubricin  chains can adsorb on both hydrophilic and hydrophobic surfaces~\cite{Chang:2009kf,Zappone:2007ev}, but their conformations in the adsorbed state strongly depend on the type of surface. These molecules can bind onto hydrophilic surfaces via their charged domains located in the central
 part of  a molecule and form  structures resembling a brush composed of linear chains. Alternatively, they can also bind onto hydrophobic surfaces with their terminal groups to form  loop brush-like structures~\cite{Chang:2009kf,Coles:2010dh,Zappone:2007ev}. Hence, based on  experimental observations and variations in the molecular conformations \textit{in vitro}, one may conclude that in reality the whole cartilage surface resembles a polymer brush composed of  linear and loop-like chains.  Although molecules mentioned so far are highly charged, and long-range interactions might be a dominant factor in the reduced frictional forces, conformation of chains -- whether they are linear or looped -- can influence inter-digitation of molecules as well as their relaxation times, and hence, friction and lubrication in tissues.

Inter-digitation effects  due to ring topology have also been  reported in the context of genetic material:  in a computational study,  unconcatenated chains of ring polymers showed a weak trend towards  inter-mixing with each other compared to mixing behaviour of linear chains in confined environments~\cite{Dorier:2013ic}.
We should also underline that ring polymers in bulk and at melt densities exhibit  different behaviour with respect to their linear counterparts~\cite{Halverson:2011fx,Cates:1986kt,Halverson:2011fc}.  The behaviour of ring polymers cannot be described by one single fractal dimension unlike linear chains in their melts: while size of  a linear chain at melt state scales as $R_{0L} \sim N^{1/2}$, a ring polymer in the melt of rings is much more compact and the characteristic size scales $R_{0R} \sim N^{1/2}$  for short chains and $R_{0R} \sim N^{1/3}$ for sufficiently large polymerization degrees, $N$. In between these two limits, an intermediate regime $R_{0R}\sim N^{2/5}$ arises~\cite{Halverson:2011fx,Cates:1986kt}.  Moreover,  even in the regimes where  the size of ring chain is  non-Gaussian, interestingly higher moments of the end-to-end distance exhibit a Gaussian-like behaviour~\cite{Halverson:2011fx}.

 Existing computational~\cite{Ou:2012cv, Spirin:2013jr,Galuschko:2010im,Carrillo:2011ic, Grest:1993uc, Liberelle:2008de,Binder:2012bp}  and  theoretical studies~\cite{Galuschko:2010im, Witten:1990tp, Semenov:1995gg, Joanny:1992tb,Klein:1996wk} on linear-polymer brushes  have shown that frictional forces acting brushes exhibit a cross-over from a linear to a non-linear regime upon increasing shear velocity.  The onset of non-linear friction typically appears at around shear rates where grafted chains begin to stretch. Ideally, one would expect a similar behaviour in the case of ring-polymer brushes. However, how the topology and peculiar scaling of ring chains alter the friction is an open question.
To the  best of our knowledge there are very few computational investigations of ring or loop polymer brushes out of equilibrium~\cite{Reith:2011co,Yin:2007ce}.  In a previous work~\cite{Yin:2007ce},  tribolgy of loop brushes near the overlap concentrations were studied, but the observed difference in frictional forces was  negligible. However, at melt densities where the correlation length -- distance between two chains -- is on order of monomer size, the situation can be different as the number of collision between brush monomers can significantly increase in a denser system.

Characterizing the systematic reduction in brush friction due to the topology of the constituting chains can improve our understanding of how nature handles friction and help the design and improvement of  advanced biomimetic lubricants.
Thus, spurred by the abundance of brush-like structures in nature and the interesting nature of ring polymers themselves in this paper we aim to study non-equilibrium behaviour of ring-polymer brushes at melt densities.  In our extensive coarse-grained MD simulations, we used neutral (uncharged) polymer brushes. Given the fact that the system we would like to mimic is under high pressure (e.g., in joints), and ionic condensation (short Debye length in physiological conditions) can effectively neutralize the chains, we believe that this approximation is reasonable  for the sake of minimizing computational cost~\cite{Spirin:2013jr} since long-range electrostatic interactions are known to be computationally expensive, particularly for dense systems. Through a detailed analysis of simulation trajectories and by employing scaling arguments,  we demonstrate that the  topology of chain in a melt brush is an important factor in the reduction of frictional forces.

In our simulations, we found that for untangled  brushes friction forces between two brushes made of linear chains are always roughly a factor of two higher than those for ring chains.  Although this difference is small, it persists for various grafting densities and chain sizes. This difference in the frictional forces of ring and linear brushes can be elucidated by the size of the overlap zone, which defines the amount of inter-digitation between two brushes. It turns out that the low tendency of ring chains to overlap with the opposing ring-brush leads to lower friction forces whereas linear chains can diffuse through the opposing brush more easily. Hence, brushes made of linear chains exhibit higher frictional forces. At very high velocities, on the other hand, segments of a ring brush take a double-stranded conformation. In turn, double-strands are less efficient in momentum transfer between the opposing brushes. The difference in friction forces between both systems is confirmed by scaling analysis.

The paper is organized as follows: First we briefly describe the simulation methodology. Next, the results obtained from non-equilibrium coarse-grained brush simulations will be discussed. By carefully analysing our data, we will relate the difference in friction forces to the intrinsic properties of ring and linear (grafted) chains. The scaling arguments for brush systems will also be discussed to infer the difference in forces along with the simulations data. We conclude the paper by the summary of our findings and future prospects.

\section{Simulation Details}

Simulations of polymer brush bilayers were performed  using  coarse-grained   “Kremer-Grest (KG)”  bead-spring model~\cite{Kremer:1990bn,Grest:1993uc}. Each individual chain of a polymer brush was composed of $N$ (or $2N$) monomers (beads) connected by bonds.  Two adjacent effective chain monomers are bonded via a   ``Finitely Extensible Nonlinear Elastic'' (FENE) potential
\begin{equation}
V^{\mbox{\tiny FENE}} (r) = -\frac 12 k r_0^2 \ln{\left[ 1 - \left(\frac
r{r_0}\right)^2 \right]},
\label{fene}
\end{equation}
where bond stiffness $k=30\,\epsilon/\sigma^2$,  distance between the beads $r= |\textbf{r}|$, and maximum bond length $r_0=1.5\,\sigma$~\cite{Kremer:1990bn}.  The interaction strength $\epsilon$ was measured in units of thermal energy $k_BT$.
The non-bonded interactions between all monomers were modeled by the
 truncated and shifted Lennard-Jones (LJ) potential
 \begin{equation}
 V^{\mbox{\tiny LJ}}(r) =\left\{ {\begin{array}{*{20}c}
   4\epsilon\left[
(\sigma/ r)^{12} - (\sigma /r)^6 + 1/4\right]  \qquad r \leq r_c \\
   0~\quad \qquad\qquad\qquad\qquad\qquad\qquad r > r_c, \\
\end{array}} \right.
%\theta(r_c-r).
\label{wca}
\end{equation}
where  $\sigma$ was chosen
as the unit of length and $r_c$ is cutoff.
For polymer-polymer interaction we have used $\epsilon=1$ and $r_c=2^{1/6}\,\sigma$.
This choice of LJ potential cutoff in combination with a monomer density of $\rho_m \approx 0.6\, \sigma^{-3}$ provides correct melt statistics~\cite{Pastorino:2007hg,Galuschko:2010im,Spirin:2013jr}.

Polymer brushes composed of chains with either linear or ring-like topology were studied. We performed simulations with linear-polymer brushes composed of $N=60$, $100$  monomers, and ring-polymer brushes of $N=120$, $200$ monomers  per grafted chain. Chains were grafted on a square-lattice surface with  dimensions  $42\,\sigma\times 36\,\sigma$.  The ring chains were grafted by one of their monomers on the surface. No equation of motion was solved for surface monomers (surface monomers were immobile during simulations).  The surface and the brush monomers  interact via Eq.~(\ref{fene}), and Eq.~(\ref{wca}). The grafting densities of linear chains  are $\rho_g^L=0.11\,\sigma^{-2}, 0.25\, \sigma^{-2}$, which respectively  corresponds to inter-anchored monomer distances  of $3\,\sigma\,, 2 \sigma$ at the surface. In the case of ring brushes, to obtain an equal monomeric density, $\rho_m \approx 0.6\;\sigma^{-3}$, to the linear-brushes, the grafting densities were taken as $\rho_g^R = 0.5 \rho_g^L$ at the same inter-plate distance $D$. Note that $D$ and $\rho_g^{L,R}$ are connected via $\rho_m \approx N \rho_g^{L,R}/D$.
In order to construct brush bilayer systems, first two non-interacting single brushes were generated as mirror images of one another. While one of the brushes is fixed at $z=0$, the other brush is brought into contact slowly at the desired inter-plate distance $z=D$ to obtain the same monomer density for each $\rho_g^{L,R}$ and  $N$. Finally, the systems are run at $v=0$ for at least $10^7$ MD steps to allow chains to relax.

The molecular dynamics  simulations were performed by solving the Langevin equation of motion, which describes the Brownian motion of a set of interacting monomers, as
\begin{equation}
m\ddot{\mathbf r}_i =   \mathbf F_i^{\mbox{\tiny
LJ}} + \mathbf F_i^{\mbox{\tiny FENE}} -\Gamma\dot{\mathbf r}_i + \mathbf F_i^{\mbox{\tiny R}}, \,\,\,
i=1,\ldots,{ N},
\label{langevin}
\end{equation}
where  $\mathbf r_i=[x_i,y_i,z_i]$ is the position of $i$--th monomer. $\mathbf F_i^{\mbox{\tiny LJ}}$ and $\mathbf F_i^{\mbox{\tiny FENE}}$ in Eq.~(\ref{langevin}) are respectively  LJ and FENE forces exerted on the $i$--th monomer and given by derivatives of Eqs.~(\ref{fene}) and~(\ref{wca}) with respect to $\mathbf r_i$.
The effect of the implicit solvent in Eq.~(\ref{langevin})  is split into a slowly evolving viscous force $-\Gamma\dot{\mathbf r}_i$ and a rapidly fluctuating stochastic force $\mathbf
F_i^{\mbox{\tiny R}}$. This random force $\mathbf
F_i^{\mbox{\tiny R}}$ is related to the friction coefficient $\Gamma$ by the fluctuation-dissipation
theorem $\langle \mathbf F_i^{\mbox{\tiny R}}(t) \mathbf F_j^{\mbox{\tiny R}}(t')\rangle = k_BT\Gamma \delta_{ij}\delta(t-t')$. The friction coefficient used in simulations was $\Gamma=0.5\,m\tau^{-1}$, where $m=1$ is the monomer mass and time was measured in units of $\tau =\sqrt{m\sigma^2/\epsilon}$.
The integration step was taken to be $\Delta \tau = 0.002\,\tau$. The velocity Verlet  scheme was used for numerical integration of equations of motion
Eq.~(\ref{langevin}).
All simulations were performed in the $N_mVT$-ensemble, i.e., constant volume $V$,  total particle number $N_m$ and temperature $T$. The system temperature was set to the value $T=1.68~\epsilon/k_B$ with $k_B=1$ \cite{Pastorino:2007hg}.
Periodic boundary conditions were introduced in the lateral directions, i.e. in $\hat{x}$ and $\hat{y}$ whereas
in the $\hat{z}$-direction fixed boundary conditions were imposed.
Simulations were carried out using molecular package LAMMPS~\cite{Plimpton:1995wla}.
Simulation snapshots were rendered using Visual Molecular Dynamics (VMD) \cite{vmd}.
\begin{figure}
\includegraphics[ width=8cm,viewport= 0 0 600 360, clip]{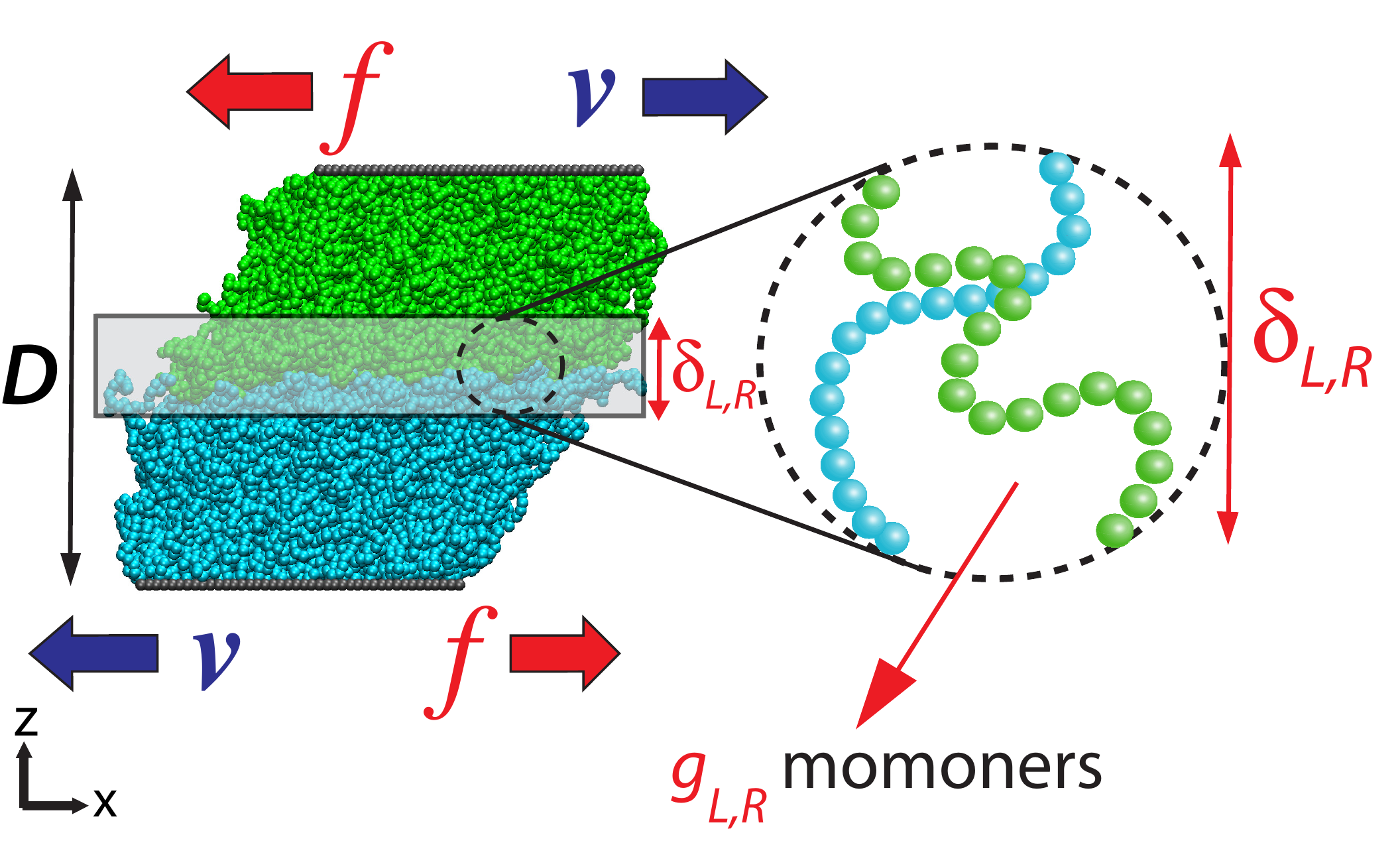}
\caption{Scheme of brush bilayer shear simulation. The force $f$ acts in the direction opposite  to velocity $v$. Chains at the top and bottom brush are rendered in different colors. The overlap zone (OZ) with thickness  $\delta_{L,R}$ is indicated by a milky-white region. The dashed circle shows the segments of chains with $g_{L,R}$ monomers inside the penetration zone.  Here $N=60$, $\rho_g^L=0.25\; \sigma^{-2}$ and $D=50\;\sigma$. % and $ v=0.125\;\sigma/\tau$.
%Images are obtained via VMD.
}
\label{fig:illustration1}
\end{figure}

 The non-equilibrium shear simulations were performed as shown in Fig.~\ref{fig:illustration1}. Both plates grafted by polymers were moved laterally in the opposite $(\pm \hat{x}$-directions) at prescribed velocities in a range  of $v \approx 10^{-3}\, \sigma/\tau$ and  $v \approx 1\, \sigma/\tau$. The inter-plane distance $D$ was kept constant  during shearing.  For each plate velocity, all system were run for $10^7$  MD steps until the steady state in friction force was reached, i.e. error bars in  time-averaged quantities were independent of simulation time. Error bars were calculated via block-averaging. For the purpose of data analysis, additional simulations were run for $10^6$ up to $10^8$ MD simulation steps depending on the velocity and in order to obtain proper plate displacements.  As a rule each plate was displaced by at least 5 times in the corresponding $\pm \hat{x}$-directions.
 To avoid any bias on friction forces or vertical chain diffusion while the system was sheared, the thermostat was applied only in the $\hat{y}$-direction~\cite{Pastorino:2007hg,Ou:2012cv}.

In this paper, we report frictional forces that are the forces acting on the plates to keep them at a constant velocities,  as well as averaged brush properties that are calculated from monomer trajectories by employing in-house analysis programs. Unless otherwise noted, all results presented in this paper were averaged over time.

\section{Results and Discussion}
\begin{figure}
\begin{subfigure}{0.95\textwidth}
\includegraphics[ width=\textwidth, viewport= 0 0 800 840, clip]{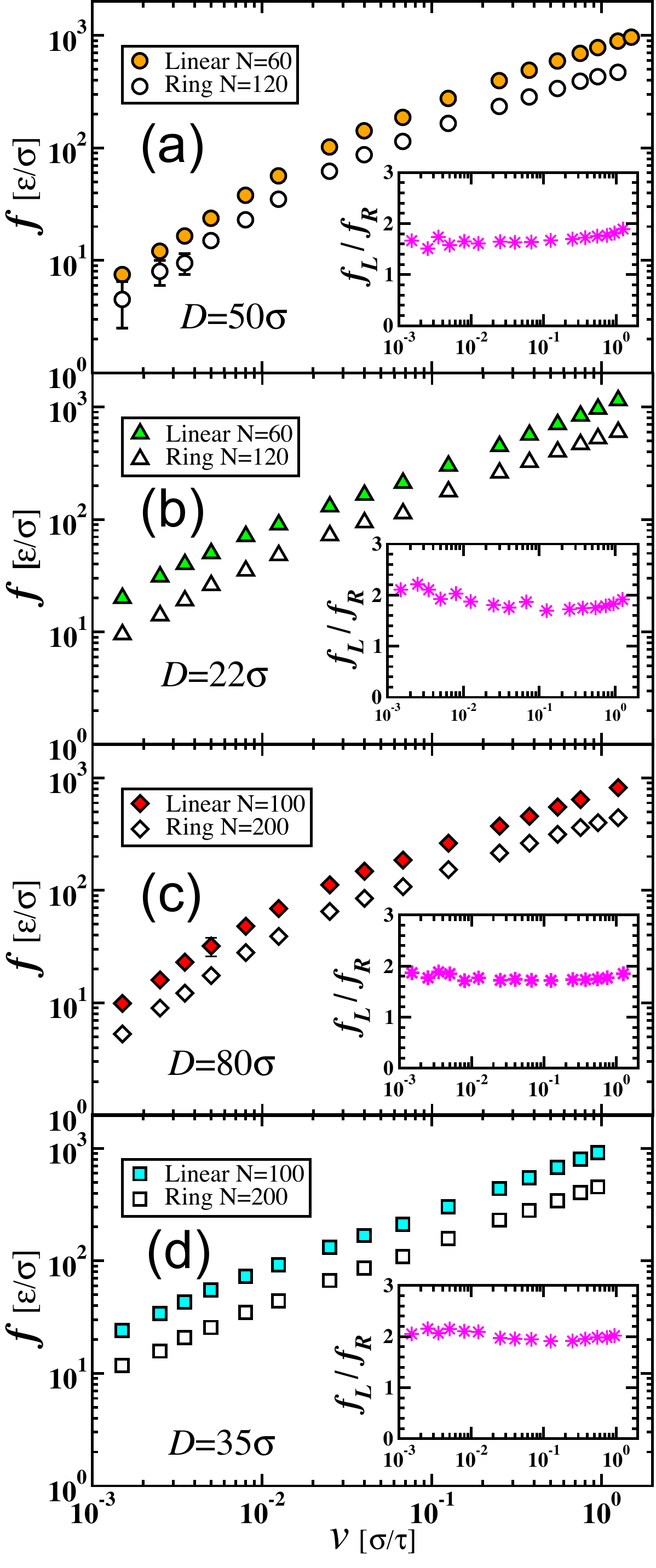}
\end{subfigure}
\caption{ Total friction forces $f$  as a function of plate  velocity $v$ for linear and ring brushes. The grafting densities are
a) $\rho_g^L=0.25 \sigma^{-2}$ b) $\rho_g^L=0.11 \sigma^{-2}$ c)  $\rho_g^L=0.25 \sigma^{-2}$ d)  $\rho_g^L=0.11 \sigma^{-2}$  with $\rho_g^R = 0.5 \rho_g^L $ for all plots.}
\label{fig:forces}
\end{figure}

In the course of  simulations,  two compressed polymer brushes were moved in the opposite ($\pm \hat{x}$) directions by driving  both planes grafted by chains at prescribed velocities $v$ as shown in Fig~\ref{fig:illustration1}. Each plane experiences  a friction force in the opposite direction to its motion due  to relative motion of brushes.
In Fig.~\ref{fig:forces}, the friction force for linear brushes $f_L$ (filled symbols) and for ring brushes $f_R$ (open symbols) are shown as a function of velocity, $v$, for various brush systems with different polymerization degrees  $N$ and chain topologies. The overall velocity dependence of friction forces is consistent with previously reported results~\cite{Galuschko:2010im,Spirin:2013jr,Goujon:2012dh,Carrillo:2011ic}: The frictional force increases linearly up to a threshold velocity, which occurs at different velocities for different brush bilayers. Above a threshold velocity, a sublinear increase occurs for all brush systems considered here.
We observe this linear-to-sublinear transition in most of our systems except the case with inter-plate distance $D=35\, \sigma$, where only the onset of the non-linear regime can be seen in Fig.~\ref{fig:forces}d.  As  we will discuss in  more details in the following sections, this is due to the fact that the threshold velocity separating the linear and non-linear force regimes depends on segment size inside the overlap volume, where two opposing brushes can co-exist.  Hence, as the size of the average segments  increases inside the overlap volume, much slower velocities are required to observe the linear-force regime.
The comparison of friction forces acting on linear and ring brushes, which is the main motivation of this work,  shows that friction forces for linear brushes are always higher than those acting on ring brushes, i.e. $f_L> f_R$. The numerical value of  the ratio between these two forces $f_L/f_R \approx 2$ as can be seen in the insets of Fig.~\ref{fig:forces}. The ratio holds for a broad range of shearing velocities $(v \approx 10^{-3}$-$10^0\, \sigma/\tau$) used in simulations (see the insets of  Fig.~\ref{fig:forces}).

Average normal forces acting on the plates in the $\hat{z}$-direction were also measured: We observed that normal pressures exhibit a weak dependence on velocity, i.e. a three order of magnitude increase in velocity leads to a roughly  $10\%$ increase in the normal forces for both linear and ring brushes  (data not shown). At high velocities, a slight decrease in all measured pressures is observed. Similar to the frictional forces, the normal pressures measured for linear brushes are factor of two higher than ring brushes, $p_z^L/p_z^R \approx 2$. However, if total normal forces are rescaled by the number of grafted chains for each system, normal forces per chain are of almost equivalent in both systems.

In what follows,  we will discuss the observed difference in the frictional forces acting on linear and ring brushes. We will consider  linear and non-linear force regimes separately and demonstrate that the ratio $f_L/f_R \approx 2$ is  related to the topology of chains and of their velocity-dependent conformations.

\subsubsection{Linear regime}
In this subsection we will consider the regime where  frictional forces  increase linearly with the driving velocity. %. as shown in Fig.~\ref{fig:forces}.
If two dense polymer brushes are brought into contact at the inter-plate distance $D>R_0$,  as shown in Fig~\ref{fig:illustration1}, where $R_0$ is  equilibrium size of a free chain in bulk, chain  segments near the free ends of grafted chains can interpenetrate through the opposing brush. Hence, the overlap zone (OZ) where monomers of opposing brushes can mutually interact can be defined (see a rectangular region in Fig.~\ref{fig:illustration1}).

From simulation trajectories, the thickness of the OZ can be obtained using the cross product of  monomer density profiles of top $\rho_{\mbox{\tiny top}}(z)$ and bottom $\rho_{\mbox{\tiny bott}}(z)$ brushes.  The cross product is non-zero only if monomers from  both parts of brush coexist at the same $z$ coordinate. The width of the OZ can be calculated from the cross-product-weighted averages as
\begin{equation}
\centering
\delta (v)  \equiv 2 \left[ \int_{-D/2}^{D/2 }z^2 \omega (z,v) dz - \left(\int_{-D/2}^{D/2 } z  \omega (z,v) dz \right)^2 \right],
\label{eq:delta_std}
\end{equation}
where we defined the cross product  as $\omega (z,v) \equiv \rho_{\mbox{\tiny top}}(z,v)\rho_{\mbox{\tiny bott}}(z,v)$.  Hence, from Eq.~(\ref{eq:delta_std}), equilibrium value of OZ thickness reads $\delta(v \rightarrow 0) \equiv \delta_0$. Note that in this paper we use  subscript ``zero" to indicate equilibrium (or linear-response) values of corresponding parameters (i.e. plate velocity $v=0$) whereas subscripts
\textit{L}  and \textit{R}  refer respectively to linear and ring brushes.
\begin{figure}
\centering
\includegraphics[width=8cm,viewport= 0 0 650 550]{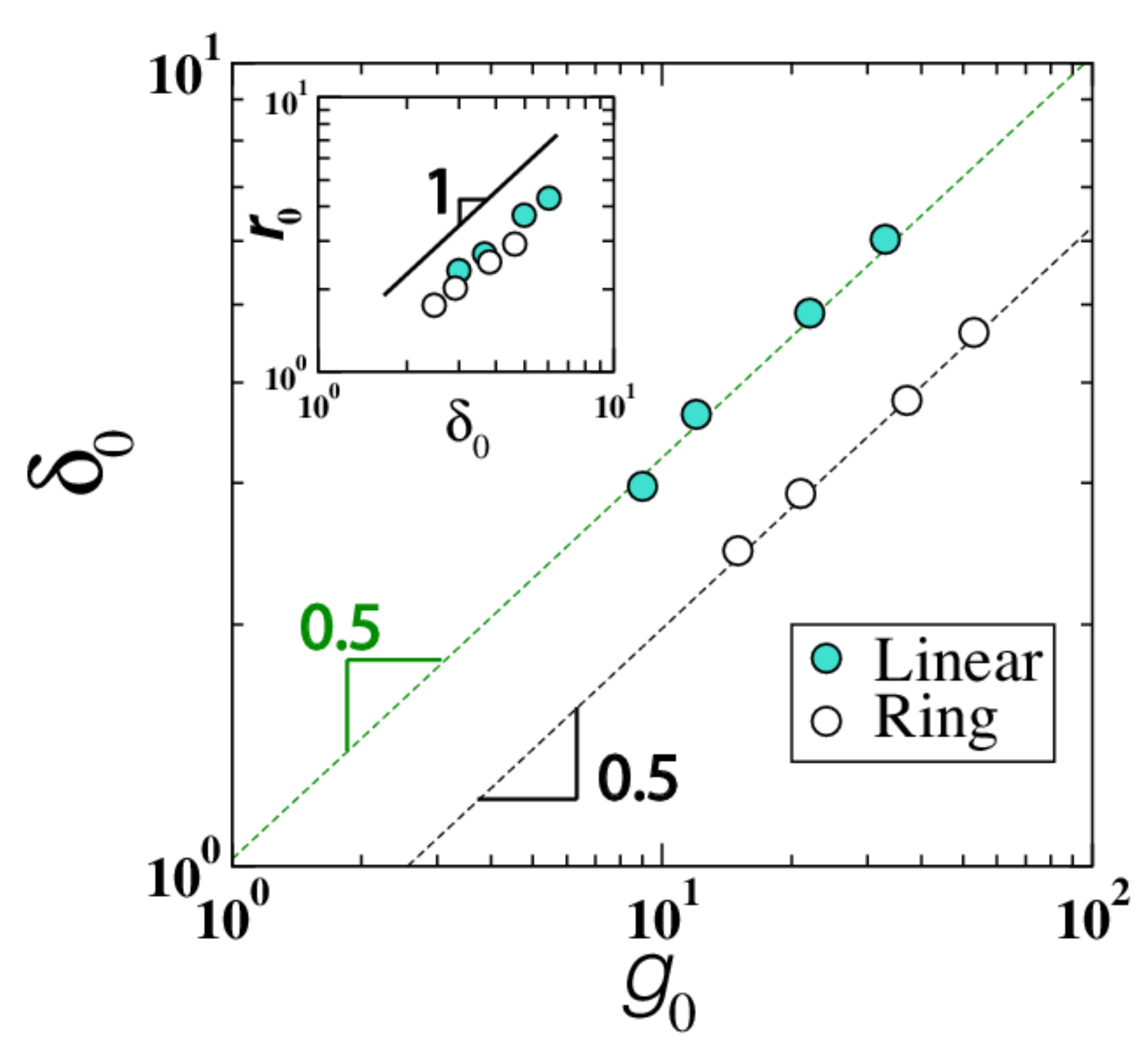}
\caption{The equilibrium values of the width of the overlap zone (OZ) versus number of monomers $g_0$ per segment of a chain located inside the OZ  for
linear (full symbols) and ring (empty symbols) brushes. Dashed lines represent fitted power laws: $\delta_0\approx1.0 g_0^{1/2}$ (for linear chains) and $\approx0.625 g_0^{1/2}$ (for rings).
The inset shows fluctuation of the end-to-end size of segments $r_0\equiv \langle r_0^2\rangle^{1/2}$ inside the OZ  as a function of $\delta_0$.
Solid line stands for fitted scaling $r_0\propto \delta_0$.}
\label{fig:delta_vs_g}
\end{figure}

The equilibrium values of the widths of the OZ for linear and ring brushes $\delta_{0L,0R}$ are shown in Fig.~\ref{fig:delta_vs_g} as a function of
 the number of monomers $g_{0L,0R}$ per
segment of a chain inside the OZ for linear and ring
brushes.
As a rule a segment is considered as being present inside the OZ if $z$ coordinate of the $N$-th  monomer of the corresponding chain satisfies the
  following condition: $D/2  -  \delta_{L,R}< z < D/2 + \delta_{L,R}$.
As we perform our simulations nearly at melt density $(\rho_m  \approx 0.6\,\sigma^{-3})$  the random-walk statistics  allows us to express
square-root of the end-to-end distance
$r_{0L,0R} \equiv  \langle r_{0L,0R}^2 \rangle^{1/2}$ of a segment inside the OZ and  composed of  $g_{0L,0R}$ monomers as $r_{0L,0R}  \sim g_{0L,0R}^{1/2}$.  The latter
 expression is valid  for short segments.
 If the size of the segment determines the width of the OZ, then the  relation   $r_{0L,0R} \approx \delta_{0L,0R}$ should also hold. Indeed, this is what we observe in Fig.~\ref{fig:delta_vs_g} and its inset for both linear and ring brushes:  $r_{0L,0R}  \approx \delta_{0L,0R} \sim g_{0L,0R}^{1/2}$.    %even for the ring brushes for the segment sizes considered here.
Interestingly, we found that at the same inter-plate distance $D$  grafted chain in ring brush can occupy the overlap volume with more monomers than a chain in linear brush $(g_R>g_L)$.   This finding demonstrate that inside the OZ segments of ring brushes are more compact as compared to segments of linear brushes.  This fact is also supported
 by Fig.~\ref{fig:delta_vs_g} since the pre-factor of the power-law fit is 1.6 times larger for linear brushes. The compactness of segments in ring brushes yields to a narrower OZ.
As a remark,  for very long  segments in ring brushes  $(g_R \gg 1)$ the width of the OZ should in the asymptotic limit converge to $ \langle r_{0R}^2 \rangle^{1/2}  \approx \delta_{0R} \sim g_{0R}^{1/3}$. In that case:  $\delta_{0L} /\delta_{0R} \sim g_{0L}^{1/2} /g_{0R}^{1/3}$.

%At equilibrium, i.e., at $v=0$, in the scaling level, the width of the OZ between two dense brushes scales as~\cite{Witten:1990tp,Milner:1988ud,Zhulina:1991vl}.
%
%\begin{equation}
%\centering
%\delta_{0L,0R} \approx \left( \frac{N_{L,R}^2 b^4}{D}\right)^{1/3}\approx \left( \frac{R_{0L,0R}^4}{D}\right)^{1/3},
%\label{eq:delta0}
%\end{equation}
%
%where $b$ is effective monomer size.

 Since two opposing brushes can only interact inside the OZ, the observed frictional forces should be related to the width of OZ.  The grafted chains  are not static as they constantly diffuse in and out of the overlap volume which is given by $A \times \delta_{L,R}$ where $A$ is the area of a grafting plate due to thermal fluctuations. When two plates are moved at  $v>0$, any chain segment that enters the OZ  feels a  flow induced by the relative motion of monomers moving in the opposite direction.  The force acting on each monomer inside the overlap volume can be  most generally expressed via Stokes drag $f_{m} \approx \zeta v$, where $\zeta$ is the monomeric friction coefficient.  The total friction force acting on $\rho_m A \delta$ monomers inside the overlap volume can be expressed for both linear and ring brushes as follows
\begin{eqnarray}
f_{L,R} \approx  \zeta_{0} v \rho_m \delta_{L,R} A  \Omega_{L,R}.
\label{eq:totalforce}
\end{eqnarray}
In Eq.~(\ref{eq:totalforce}), we defined  $\Omega_{L,R}$ which quantifies the number of binary collisions between the monomers of two opposing brushes. As we will see shortly,  although there are $A \times \delta$ monomers inside the overlap volume, only  a fraction of them participate in momentum exchange between opposing brushes. Hence,  $\Omega_{L,R}$ is an important quantity to distinguish the friction among different brush systems.
In addition, the equilibrium value of the monomeric friction coefficient $\zeta_0$ is equal for monomers of both linear and ring brushes.
This is due to the fact that on the time scales comparable with time between inter-bead collisions monomer cannot know whether it belongs to a linear or ring chain.

\begin{figure}
\includegraphics[width=8cm, viewport= 0 0  500 600, clip]{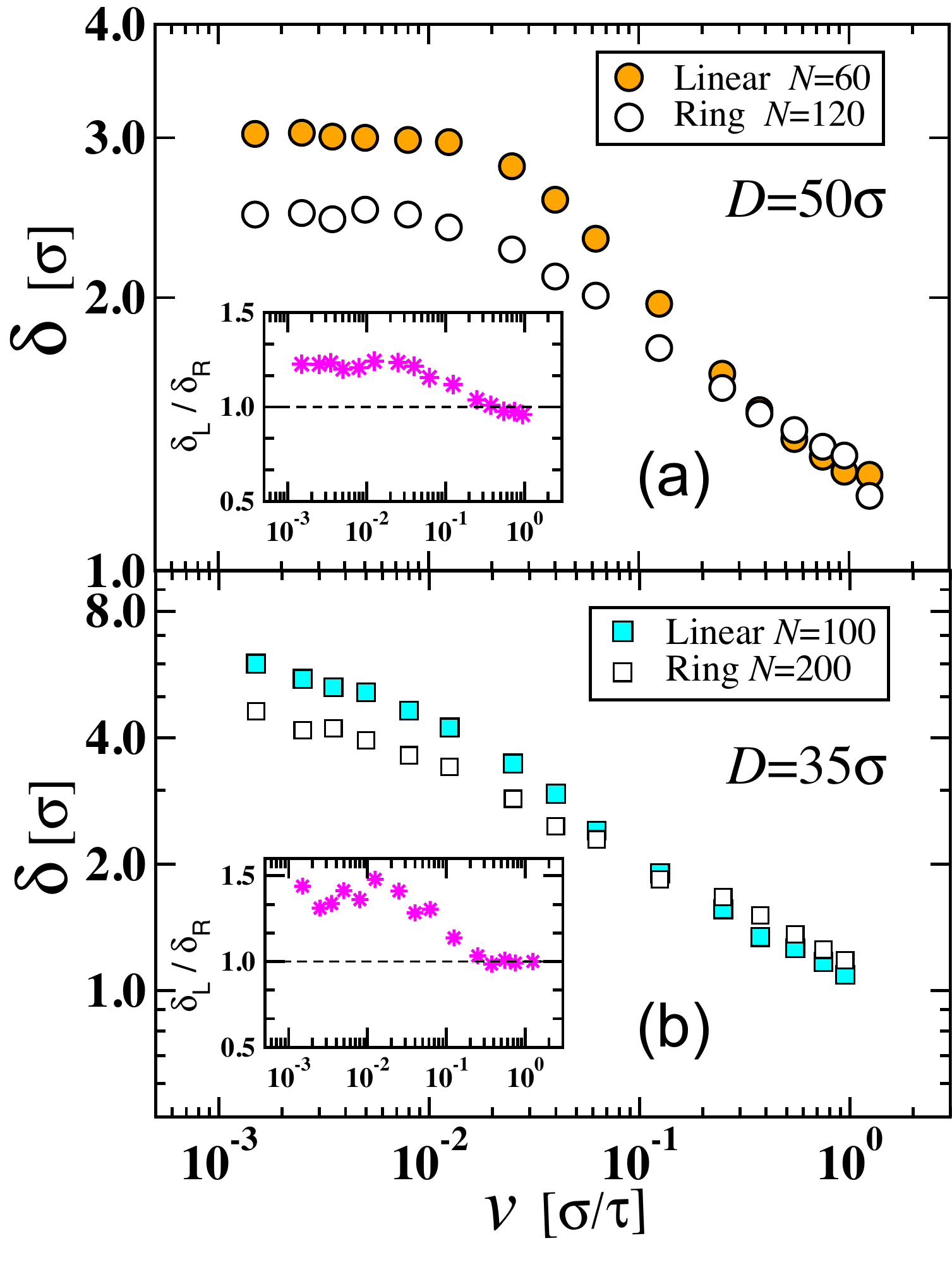}
\caption{ The nonequilibrium thickness $\delta$ of the overlap zone for linear and ring brushes plotted  as a function of plate velocity $v$. The values of $\delta$ were calculated using Eq.~(\ref{eq:delta_std}).
The insets show the ratios.
The grafting densities are a) $\rho_g^L=0.25 \sigma^{-2}$ and b)  $\rho_g^L=0.11 \sigma^{-2}$  with $\rho_g^R = 0.5 \rho_g^L $ for all plots. }
\label{fig:penzone}
\end{figure}

Based on the linear-response theory, which states that any quantity takes its equilibrium value if the perturbation is small, we replace the quantities appearing in Eq.~(\ref{eq:totalforce}) with their equilibrium values as $v \rightarrow 0$, i.e., $\delta_L \approx \delta_{0L}$ and  $\delta_R \approx \delta_{0R}$ and similarly $\Omega_L \approx \Omega_{0L}$ and  $\Omega_R \approx \Omega_{0R}$. Thus, the frictional force ratio in the linear-force regime reads
\begin{eqnarray}
\frac{f_L}{f_R}  \approx   \frac{ \delta_{0L} \Omega_{0L} } {\delta_{0R} \Omega_{0R} }.
\label{eq:ratio1}
\end{eqnarray}
The validity of Eq.~(\ref{eq:ratio1}) can be verified from the simulation data by calculating  the equilibrium values of $\delta_{L,R}$ and $\Omega_{L,R}$ for both linear or ring brushes.
From now on, to keep the manuscript more compact we will only show data from four of our eight brush systems. All comparisons and discussions are valid for the systems which are not shown here as well.
We show our non-equilibrium simulation results for the thickness of OZ $\delta_{L,R}$ as a function of plate velocity $v$ for various linear and ring brushes in Fig.~\ref{fig:penzone}. Each frame of Fig.~\ref{fig:penzone} compares the OZ width  of ring-brushes  (open symbols) to  those obtained for linear brushes (filled symbols) at the same inter-plate distance $D$. Independently of the chain topology  the  width of the OZs exhibit a plateau for all brush systems as $v \rightarrow 0$. The plateau values of the OZs, which are $\delta_L \approx \delta_{0L}$ and  $\delta_R \approx \delta_{0R}$, persist up to roughly the plate velocities  where the linear-force regime observed in Fig.~\ref{fig:forces} also ends.
Indeed, as we proposed for Eq.~\ref{eq:ratio1}, the widths of OZs $\delta_{L,R}$  are equal to their equilibrium values. This is due to the fact that within the linear-force regime, for which the frictional forces change linearly with velocity,
 values of $\delta_{0L,0R}$  are nothing but the size fluctuations in the $\hat{z}$-direction of the chain segments inside the OZ.  As also illustrated in Fig.~\ref{fig:illustration1} the size of a segment with $g_{L,R}$ monomers defines the width of the OZ.

\begin{figure}
\includegraphics[width=8cm, viewport= 0 0 500 500, clip]{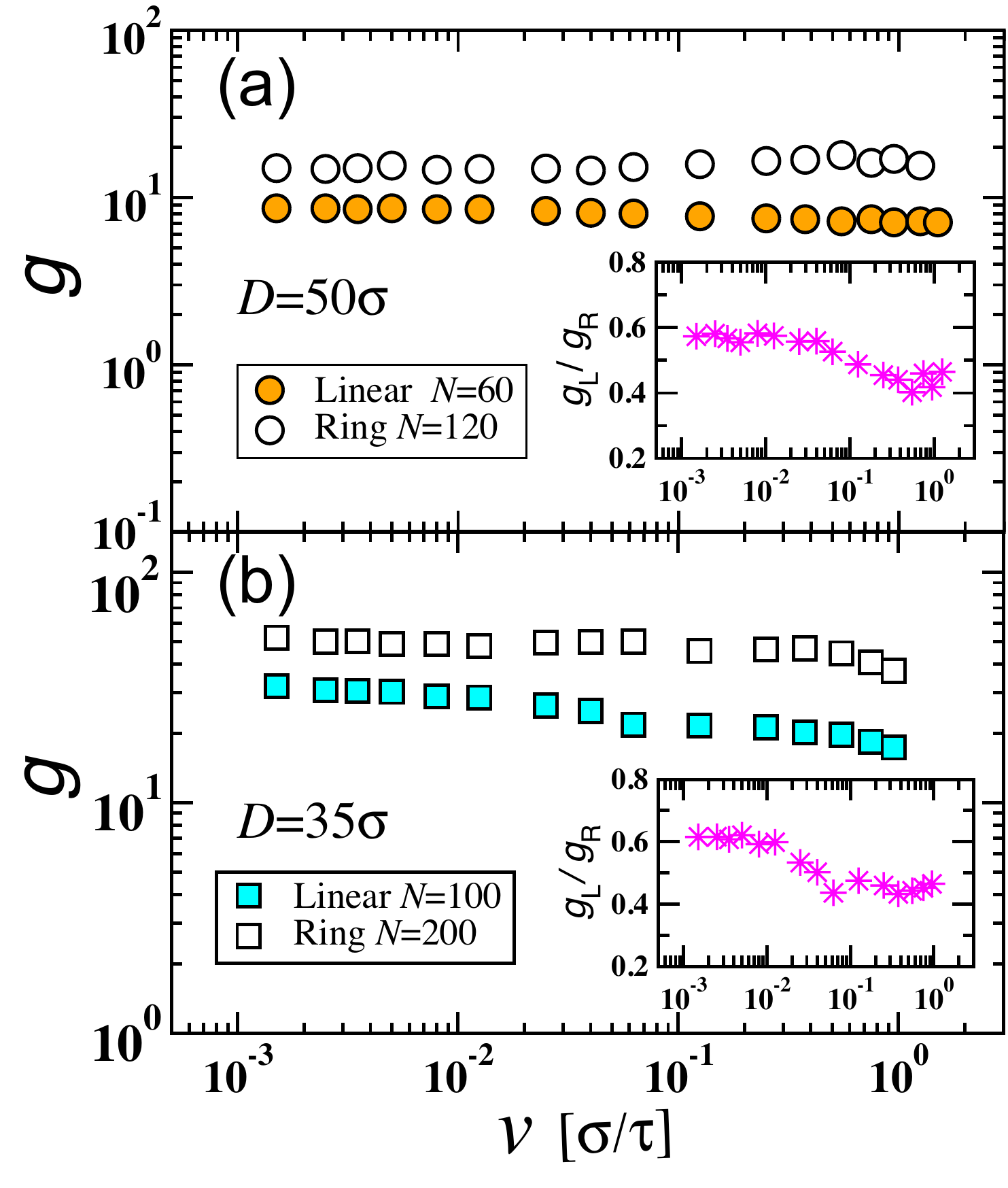}
\caption{The number of monomers $g$ inside the overlap zone per participating chain for ring and linear brush bilayers as a function of the plate velocity $v$. Insets show the ratios. The grafting densities are a) $\rho_g^L=0.25 \sigma^{-2}$ and b)  $\rho_g^L=0.11 \sigma^{-2}$  with $\rho_g^R = 0.5 \rho_g^L $ for all plots. }
\label{fig:g}
\end{figure}

The number of monomers per segment $g_{L,R}$ inside the OZ for linear and ring brushes was analyzed and is shown in Fig.~\ref{fig:g} as a function of the plate velocity $v$ using the same color code also as in Figs.~\ref{fig:forces} and~\ref{fig:penzone}.  In the entire velocity range considered here the following ratio holds  $g_R \approx 2 g_L$ for the same value of $D$. At slow velocities  $(v \rightarrow 0)$ the ratio of monomers per segment $g_{0L}/g_{0R}$ is  $\approx 0.6$. The ratio drops to $g_L/g_R \approx 0.4$ at higher velocities $(v>10^{-1}\, \sigma/\tau)$.

\begin{figure}
\includegraphics[width=8cm, viewport= 0 0  500 560, clip]{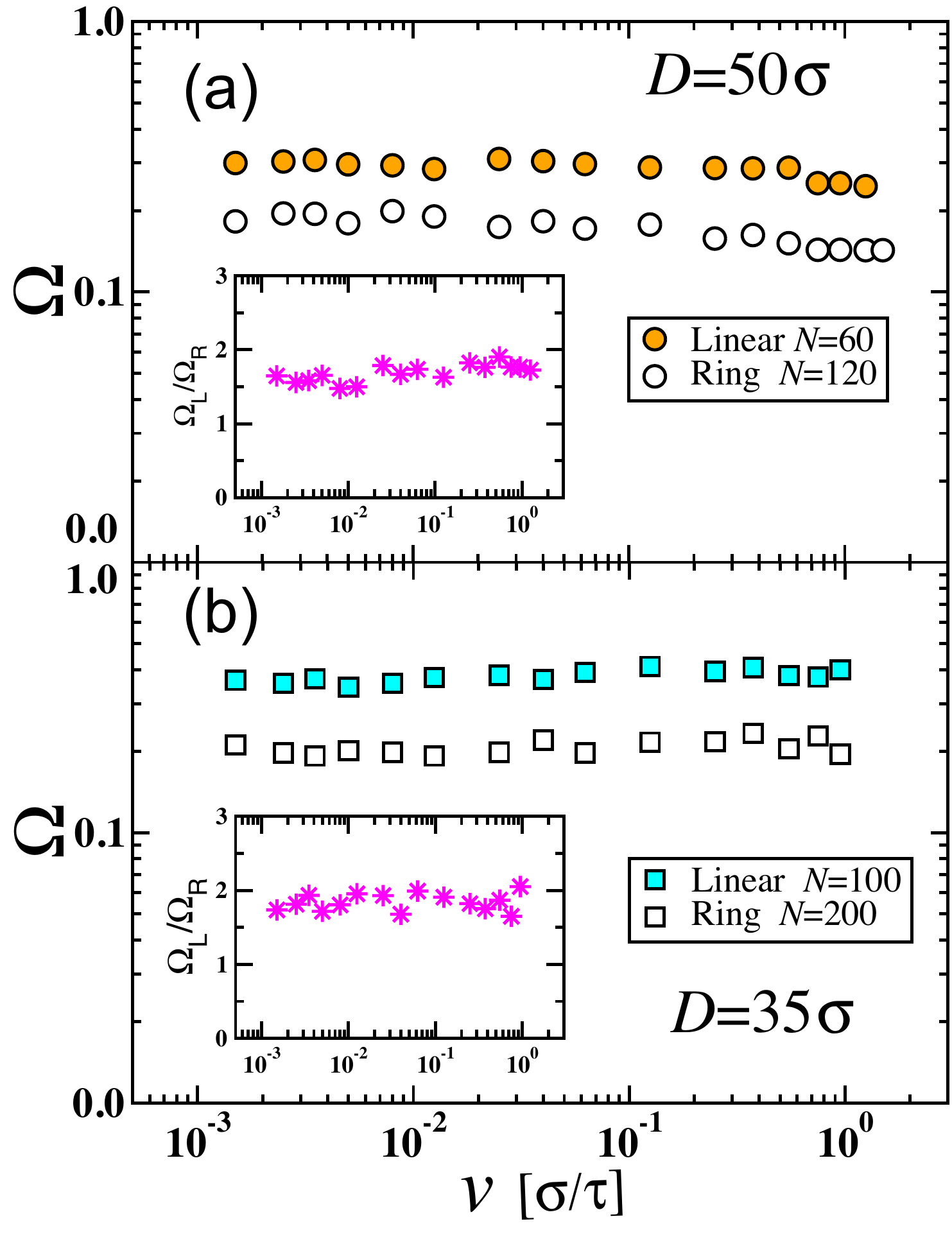}
\caption{The binary-collision factor $\Omega$ inside the overlap zone between two brushes as a function of the plate velocity $v$. See the text for details. Insets show the ratio of linear-brush and ring-brush data. The grafting densities are a) $\rho_g^L=0.25 \sigma^{-2}$ and b)  $\rho_g^L=0.11 \sigma^{-2}$  with $\rho_g^R = 0.5 \rho_g^L $ for all plots.}
\label{fig:omega}
\end{figure}

Similar trajectory analysis was also conducted to determine the number of binary collisions $\Omega_{L,R}$ inside the overlap volume: to count collisions, distances between the monomers from opposing brushes  are calculated. If the distance is equal or smaller than the  cutoff distance of LJ potential defined in Eq.~(\ref{wca}),  this specific pair is counted as a colliding pair.  To obtain $\Omega_{L,R}$ and compare the number of collisions for ring and linear chains, the total number of collisions is rescaled by the corresponding  value of $\rho_m A \delta_{L,R}$. In Fig.~\ref{fig:omega}, we show the binary-collision fractions $\Omega_{L,R}$, as well as their ratios in the insets. Interestingly, for linear brushes there are more  collisions as compared to ring brushes, i.e. $\Omega_L > \Omega_R$ for the entire velocity range although $g_R \approx 2 g_L$.
An average ratio of  $g_L/g_R \approx 0.5$ indicate that even though a single chain in ring brushes  can occupy the overlap volume with more monomers than a chain in linear brushes  ($g_R>g_L$)  the compactness of the segments in ring brushes  yield to a narrower OZ. This is demonstrated in Fig.~\ref{fig:delta_vs_g} where the pre-factor of the slope is bigger for linear brushes. The compactness of segments in ring brushes also results in a smaller number of collisions between the monomers of opposing brushes. Hence,  $f_L/f_R \approx 2$.
In addition, as expected values of $\Omega_L$ and $\Omega_R$ are close to their  values  as $v \rightarrow 0$, which confirms that $\Omega_L \approx \Omega_{L0}$ and $\Omega_R \approx \Omega_{0R}$.

At the same plate separation $D$, the ratios of  plateau values  are $\delta_{0L} / \delta_{0R} \approx 1.3 \pm 0.05$ from Fig.~\ref{fig:penzone}  and $\Omega_{0L} / \Omega_{0R} \approx 1.7 \pm 0.1$ from Fig.~\ref{fig:omega}.   Plugging the later expression  into  Eq.~\ref{eq:ratio1} gives the ratio of friction force  for the linear-force regime $f_L / f_R  \approx    2.3$, which is consistent with the ratios of forces obtained in the insets of Fig.~\ref{fig:forces}).

 As the frictional forces  increase linearly (see Fig.~\ref{fig:forces}) in principle none of terms in Eq.~(\ref{eq:totalforce}) should have any velocity dependence. Indeed, this is what one would expect at the linear-response level: at slow enough velocities ($<10^{-1}\,\sigma/\tau$)  the conformation of chain segments within the OZ  are not affected by the velocity. As a result, the ratios of forces for linear and ring brushes are constant and given by Eq.~(\ref{eq:ratio1}).

Our analyzes confirm that in ring-brush bilayers, the amount of inter-digitation of chains is weaker compared to that in linear-brush bilayers at slow driving velocities, where the friction increases linearly with velocity.

\subsubsection{Non-linear regime}

In this subsection we will focus on the regime where frictional forces demonstrate sublinear plate velocity $v$ dependence.
As can be seen in Fig.~\ref{fig:forces} nonlinear regime is observe for both linear and ring bilayers.  The indication of the non-linear regime is the shrinkage of overlap volume  with increasing velocity  $v$ as shown in Fig.~\ref{fig:penzone}. Onset of the non-linear regime occurs at a threshold velocity $v=v^*$. Indeed, $v^*$  has different values for different chain sizes as can be noticed in Figs.~\ref{fig:forces} and~\ref{fig:penzone}. Before we further continue our discussion on the non-linear regime, we will briefly discuss two scaling predictions for the threshold velocity, and refer them as $v_I^*$ and $v_{II}^*$.

We will temporarily drop the indexes for ring and linear brushes  for simplicity and refer $g_0\equiv g_{0L}\equiv g_{0R}$ and $\delta_0\equiv \delta_{0L}\equiv \delta_{0R}$, etc.  If a segment with $g_0$ monomers enters the OZ, the force acting on this segment in the linear regime can be expressed  as $f_{ch}  \approx \zeta_0 g_0 v$.  As we have discussed in the previous section, below the threshold velocity  the chains are still Gaussian. Hence, as expected from a Gaussian chain, segment sizes in  $\hat{x}$, $\hat{y}$ and $\hat{z}$-directions are not coupled, i.e. fluctuations in size in all direction are independent of each other. However, as the velocity is increased, the frictional force per segment inside the OZ reach a threshold force $f_{ch}^*  \approx k_B T/b   \approx \zeta_0 g_0 v^*_I$ at $v=v^*_I$~\cite{Semenov:1995gg}. The force on the segment given by $f_{ch}^* \approx  k_B T/b$ is high enough to align individual bonds in the direction of the relative motion. Thus, vertical segment fluctuations in the $\hat{z}$-direction, which define the width of OZ, will be affected by the shear in the acting in $\hat{x}$-direction.  As a result the OZ decreases with respect to its equilibrium value $\delta< \delta_0$ at $v>v_I^*\approx k_B T/b \zeta_0 g_0 \sim 1/g_0$~\cite{Semenov:1995gg}.
On the other hand, in a concentrated solution one may argue that  the threshold velocity occurs at slower velocities and is given by $f_{ch}^{*} \approx  \zeta_0 g v_{II}^* \approx k_B T/\delta_0$ that leads to   $v_{II}^* \approx k_B T/b \zeta_0 g_{L/R}^{1+\nu}$~\cite{Galuschko:2010im}, where $\nu$ is the scaling exponent ($\nu=1/2$ for an ideal chain and $\nu=0.588$ for a swollen chain in 3D).  In our simulations,  hydrodynamic interactions are screened and we take  $\nu=1/2$ and obtain $v_{II}^* \sim 1/g_0^{3/2}$. The ratio of two predictions of the  threshold velocity is $v_{I}^* / v_{II}^* = g_0^{1/2}$. Unfortunately, in our simulated systems segments sizes inside the OZ are of the order of   $g_{0L,0R} \sim 10$.  Thus, we cannot capture a significant difference between the threshold-velocity predictions $v_{I}^*$ and $ v_{II}^*$. In principle, brush systems with larger  $g_{0L,0R}$ values can be designed. However, for longer segment sizes, e.g.  $g_{0L,0R} \approx 100$ chain entanglements will also come into play and introduce more complexities. Hence, throughout this work we will leave investigation of the threshold velocity $v^*$ for a separate work~\cite{self}, and refer $v^*$ only to distinguish between  the linear (any segment inside OZ is stretched less than its equilibrium size) and the non-linear force regimes (the segment end-to-end distance is more than its equilibrium size).

\begin{figure}
\includegraphics[scale=0.4]{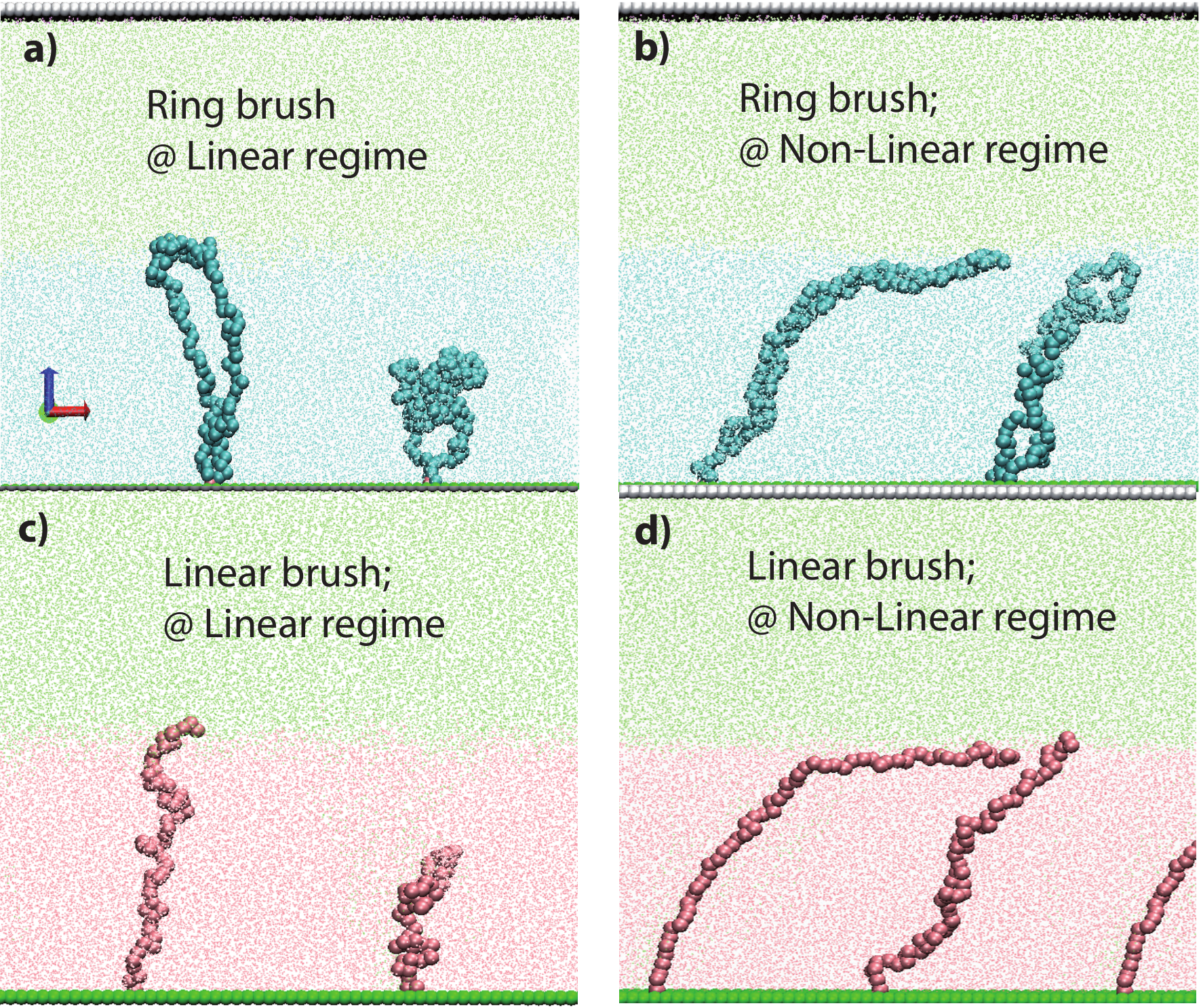}
\caption{ Snapshots of the  individual chains from ring (a and b) and linear brush bilayer (c and d) simulations in the linear-force regimes (left column) and non-linear-force regimes (right columns).  Polymerization degree of chains for linear brushes is $N=60$ whereas for ring brushes is $N=120$. Monomers of surrounding chains are rendered as points with colors referring to the top and bottom brushes.}
\label{fig:illustration2}
\end{figure}

As discussed above, at   $v> v^*$ the segments inside the OZ are not Gaussian, and are stretched due to the relative motion of brushes, as illustrated by  the snapshots given in Fig.~\ref{fig:illustration2}.  At the velocity range $v^*<v<v_{\mbox{\tiny max}}$, where the maximum velocity that we considered here $v_{\mbox{\tiny max}} \approx 1\;\sigma/\tau$, both   $\delta_L$ and $\delta_R$ decrease in a similar way  with increasing velocity as shown in Fig.~\ref{fig:penzone}. Thus,  Eq.~\ref{eq:ratio1} is still valid to explain the difference in the frictional forces.

At  around $v_{\mbox{\tiny max}} \approx 1\;\sigma/\tau$, regardless of the brush topology, the width of the OZs goes to unity $\delta_{L,R} \rightarrow \sigma$  as shown Fig.~\ref{fig:penzone}.  Hence,  at  $v \gtrsim v_{\mbox{\tiny max}}$, according to Eq.~\ref{eq:ratio1}, the ratio of friction forces for linear and ring brushes  is reduced to
\begin{eqnarray}
\frac{f_L}{f_R}  \approx  \frac{ \Omega_L } { \Omega_R }.
\label{eq:ratio2}
\end{eqnarray}
Indeed Fig.~\ref{fig:omega} shows that  the ratio of binary-collision factors for ring and linear $\Omega_{L,R}$  is even higher at high velocities and  reaches $\Omega_L  / \Omega_R \approx 2$. The limit of $\delta_{L,R} \rightarrow \sigma$  for both ring and linear brushes at high velocities implies that   there should be equal number of monomers inside the overlap volumes $\rho_m A \delta_{L,R}$. Additionally,  we know $g_R \approx 2 g_L$ from Fig.~\ref{fig:g}. This is only possible if stretched segments of ring brushes form double-stranded conformations inside the OZ as can  be seen in the snapshot given in Fig.~\ref{fig:illustration2}b. In the non-linear regime a segment of a ring brush with $g_R$ monomers inside OZ is highly stretched and adopts a double-stranded conformation. Due to double-stranding, each monomer of a segment in a ring brush has on average  three neighbours from the same segment -- two bonded and one nonbonded. This leads to smaller number of collisions with the monomers of the opposing brush.

\begin{figure}
\includegraphics[width=8cm, viewport= 0 0  500 520, clip]{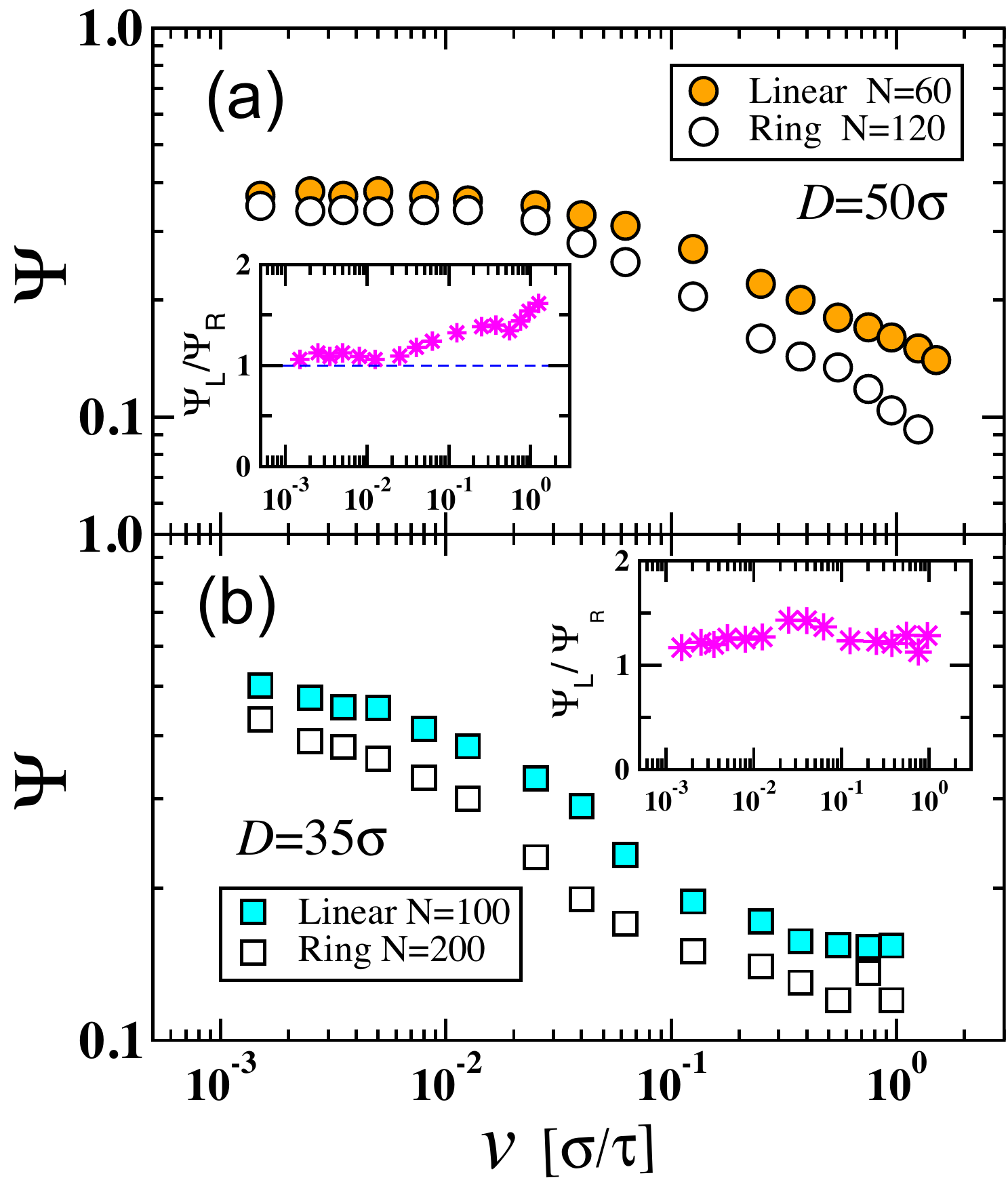}
\caption{The fraction of  participating chains inside the overlap zone as a function of the plate velocity. Insets show the ratio of linear-brush data to ring-brush data. See text for details. The grafting densities are a) $\rho_g^L=0.25 \sigma^{-2}$ and b)  $\rho_g^L=0.11 \sigma^{-2}$  with $\rho_g^R = 0.5 \rho_g^L $.}
\label{fig:psi}
\end{figure}

Another way of confirming that segments in the ring brushes are  double-stranded  inside the OZ is to calculate the fraction of ``participating''  chains,  $\Psi_{L,R}$. Since linear chains has smaller number of monomers inside the OZ $(g_L<g_R)$ larger number of chains should occupy the OZ to keep the monomer density $\rho_m$ uniform throughout the simulation box. Calculated values of $\Psi_{L,R}$  according to previously described "participating" chain description are shown in  Fig.~\ref{fig:psi} with  ratios  given in the insets. For slow velocities $(v<10^{-1}\,\sigma/\tau)$, which also corresponds to the linear-force regime, $\Psi_{0L}$ is  slightly larger than $\Psi_{0R}$ for all cases.  On the other hand at high velocities $v>v^*$ the difference between linear and ring brushes increases  and is consistent with the conditions $g_{R}>g_{L}$  and  $\delta_{L,R} \rightarrow \sigma$.
Note that in Fig.~\ref{fig:psi}b, the increase in ratio is less pronounced. Possible reason of this is our criterion for chains participating in the OZ which underestimates looping segments since we check only the  $N$th monomer of the corresponding chain within the OZ.


Finally, at ultra-high velocities  $v \gg 1\; \sigma/ \tau$, for which we cannot perform simulation since the thermostat cannot keep the system temperature uniform throughout the simulation box, one would expect that both $\delta_{L,R} \rightarrow \sigma$ and $\Omega_{L,R} \rightarrow 1$. Since the grafted chains are completely inclined and $g_{L,R} \rightarrow 1$,  two opposing brushes can only interact via few monomers near the free edges of the grafted chains. This scenario indeed yields to a high-velocity linear regime and $f_L/f_R \rightarrow 1$.

\section{Conclusion}

% Here we define $ \dot{\gamma} \equiv 2 v/\delta_{L,R}$ since only those monomers inside the overlap volume feel the relative motion of brushes.
%
%The expression for $\Psi_{L/R}$ can be obtained by considering the conversation of particle number in the OZ. The number of particles inside OZ can be written either as $\rho_ m \delta A$ or $g_{L/R} N_{ch}^{L/R} \Psi_{L/R}$. Equating these two identities gives
%%
%%
%\begin{eqnarray}
%\Psi_L \approx \frac{ N \delta_L}{g_L D},
%\label{eq:psi}
%\end{eqnarray}
%%
%%
%Here to obtain Eq.~\ref{eq:psi}, we use $\rho_m \approx N \rho_g^{L/R}/D$. In the linear regime  i.e., at $v<v^*$,  using Eq.~\ref{eq:delta0}, we can obtain  $\Psi_{0L/0R} \approx (\delta_{0L/0R}/D)^{1/2}$. Thus, Eq.~\ref{eq:totalforce} and Eq.~\ref{eq:totalforce2} match exactly in the linear regime.

In this work we demonstrated using scaling arguments and MD simulations
 that friction forces of linear chains are higher than those of ring brushes for a broad range of velocities. For slow driving velocities (or shear rates), segments of a linear brush can penetrate through opposing brush deeper compared to segments of ring brushes. This is mainly due to the compactness of the segments in the ring brushes. The compactness also lessens the number of collisions between two opposing ring brushes and results in lower frictional forces between two relatively moving brushes.  For large driving velocities (shear rates), segments inside the overlap zone between  brushes  are highly stretched regardless of the topology of chains. In the ring brushes, stretched segments form double-stranded conformations which  reduces the number of collisions with the monomers of the opposing brush.

We also observed that both friction forces and normal pressures in the linear chains are factor of two larger than those in ring brushes. This leads to equal kinetic friction coefficients $\mu= f/p$ for both systems. However, the effective viscosity $\eta_{eff}=f/v$  should be distinguishable for ring and linear chains.

 A more interesting   situation can arise when entangled brushes are considered  when chain segments can diffuse into the overlap volume with longer segments (around 100 monomers per chain for the bead-spring model). Since the bulk melts of ring chains exhibit no relaxation plateau in their stress-relaxation moduli~\cite{Kapnistos:2008dn,Halverson:2011fc}, their frictional responses should be much lower than those of entangled linear brushes. We will consider this scenario in a future publication.
\\

%\appendix
%\section{}

\begin{acknowledgments}
We thank Edward J. Banigan and Ozan S. Sariyer for their careful readings of the manuscript.\
This research was supported by the Polish Ministry of Science and Higher Education (Iuventus Plus: IP2012 005072).
% put your grant number here!!
\end{acknowledgments}

\bibliographystyle{unsrt}
\bibliography{loopbrush_jp_ae}

\end{document}